\documentclass{osa-article}
\journal{oe}

\def\al{\alpha}
\def\bt{\beta}
\def\ga{\gamma}
\def\ch{\chi}
\def\de{\delta}
\def\De{\Delta}
\def\ep{\epsilon}

\def\hb{\hbar}

\def\ka{\kappa}

\def\om{\omega}

\def\re{{\rm Re}}
\def\rh{\rho}

\def\ph{\phi}
\def\ps{\psi}

\def\si{\sigma}

\def\sinc{{\rm sinc}}
\def\ta{\tau}

\def\d{\dagger}
\def\<{\langle}
\def\>{\rangle}
\def\tint{{\textstyle\int}}
\def\tsum{{\textstyle\sum}}

\def\sb{\bar{\sigma}}
\def\d{\dagger}
\def\tr{{\rm tr}}

\def\ba{\begin{eqnarray}}
\def\ea{\end{eqnarray}}
\def\be{\begin{equation}}
\def\ee{\end{equation}}

\begin{document}

\title{Tutorial on laser linewidths}

\author{C. J. McKinstrie\authormark{1}, T. J. Stirling \authormark{2} and A. S. Helmy \authormark{3}}

\address{\authormark{1} 3 Red Fox Run, Manalapan, New Jersey 07726, USA}
\address{\authormark{2,3} Department of Electrical and Computer Engineering, University of Toronto, Toronto, Ontario M5S 3G4, Canada}

\email{\authormark{1} colin.mckinstrie@gmail.com}

\begin{abstract}
In this tutorial, the physical origins and mathematical analyses of laser linewidths are reviewed. The semi-classical model is based on an equation for the light-mode amplitude that includes random source terms, one term for each process that affects the amplitude (stimulated and spontaneous emission, stimulated absorption, and facet and material loss). Although the source terms are classical, their assigned strengths are consistent with the laws of quantum optics. Analysis of this equation shows that the laser linewidth is proportional to the sum of the (positive) source strengths for all gain and loss processes. Three-level and semiconductor lasers have broader linewidths than comparable four-level lasers, because stimulated absorption and the stimulated emission that compensates it both contribute to the linewidth.
\end{abstract}

\section{Introduction}

Masing was demonstrated in 1954 \cite{gor54} and lasing was demonstrated in 1960 \cite{mai60}. Since then, lasers have become ubiquitous in science, engineering and technology \cite{sie86,mil88,agr93,col12}. One aspect of lasers that sets them apart from other sources of light is their ability to emit waves with narrow frequency spectra. Many applications depend on this property. Hence, it is important to identify the processes that are responsible for linewidth and model them accurately.

%The Schawlow-Townes formula for the quantum-limited linewidth was an order-of-magnitude estimate based on a classical calculation by Gordon, Zeiger and Townes. MORE: Was the GZT formula BT or AT? (I guess the former.) How much should I say about the ST substitution? Their spontaneous emission argument was wrong. $kT \rightarrow hf$ is true for spectral energy density. It is not my job to do their job for them. I think less (OoME) is best.
Gordon, Zeiger and Townes \cite{gor55} modeled a maser that was driven by a beam of ammonia molecules.
By using a classical power-balance argument, they showed that the maser linewidth
\be \de\om \approx kT(\De\om)^2/P, \label{1.1} \ee
where $kT$ is the spectral power density of the reservoir that keeps the maser in equilibrium before the beam is turned on, $\De\om$ is the molecular-emission bandwidth and $P$ is the emitted power. In this formula, $\de\om$ and $\De\om$ are full widths at half maxima (FWHM).

For microwaves, $kT \gg \hb\om$, whereas for light waves, $kT \ll \hb\om$. Schawlow and Townes \cite{sch58} stated that one can apply formula (\ref{1.1}) to lasers, provided that one replaces $kT$ by $\hb\om$, which is the spectral power density that corresponds to one photon per mode. This replacement leads to the Schawlow--Townes (ST) formula
\be \de\om = \hb\om(\De\om)^2/P, \label{1.2} \ee
which is a cornerstone of laser physics.

Many papers on laser physics were written in the 1960s, by many different researchers. Some papers were based on quantum Langevin equations (stochastic differential equations for the amplitude operator), whereas others were based on density-operator equations. The calculations described in these papers are all complicated. We cite papers by Haken \cite{hak66}, Lax \cite{lax66} and Scully \cite{scu66}, which are representative of this effort and whose results are consistent. The Haken--Lax--Scully (HLS) formula for the above-threshold linewidth of a four-level laser~is
\be \de\om = \hb\om(\De\om)^2/2P, \label{1.3} \ee
where $\De\om$ is the smaller of the emission and cavity bandwidths ($1/\De\om = 1/\De\om_e + 1/\De\om_c$). Notice that formula (\ref{1.3}) differs from formula (\ref{1.2}) by a factor of 2. Measurements made by Manes \cite{man71} were consistent with a generalization of formula (\ref{1.3}) that includes the effects of inhomogeneous broadening.

The linewidth matter was settled until 1981, when Fleming \cite{fle81} measured above-threshold linewidths of semiconductor lasers, which were an order-of-magnitude larger than those predicted by the HLS formula. Henry \cite{hen82} made a first-principles analysis of spontaneous emission and showed that the HLS linewidth is increased by the factor $1 + \ep^2$, where $\ep = \ch_r^{(3)}/\ch_\imath^{(3)}$ is the ratio of the real and imaginary parts of the third-order susceptibility. (The same enhancement factor had been predicted earlier by Haug \cite{hau67}.) Because the third-order response of the material is causal, there cannot be an imaginary response (gain or loss) without a related real response (frequency shifting). This result is called the Kramers--Kronig relation \cite{jac99}. Shortly thereafter, several theoretical papers were published, which confirmed the linewidth-enhancement factor and calculated other useful quantities, such as the power and phase spectra \cite{yam83a,shi83,vah83a,vah83b,spa83,hen83}. The calculations described in these papers are also complicated. Nonetheless, their predictions were verified by experiments \cite{yam83b,har83,dai83,col83,shi84,kik84}.

In this tutorial, the physical origins and mathematical analyses of laser linewidths are reviewed. The semi-classical model described in Sec. 2 is based on a Langevin equation for the light-mode amplitude \cite{gar85,mck20a}. This equation includes random source terms, one term for each process that affects the amplitude (stimulated and spontaneous emission, stimulated absorption, and facet and material loss). Although the source terms are classical, their assigned strengths are consistent with the laws of quantum optics. Because the source terms vary randomly with time, so also does the driven amplitude. It is characterized by its two-time correlation and frequency spectrum. In Sec. 3, the amplitude equation is solved in the below-threshold regime, in which the amplitude is completely random. The solution is used to derive formulas for the temporal correlation and frequency spectrum of the complex amplitude. In Sec. 4, a simple generalization of the amplitude equation, which is called a van der Pol equation \cite{van34,jor07}, is solved in the above-threshold regime, in which the amplitude has both deterministic and random components. The solution is used to derive formulas for the temporal correlations and frequency spectra of the real and imaginary parts of the amplitude fluctuations. In Sec. 5, general formulas are derived for the below- and above-threshold linewidths, which are proportional to the sum of the source strengths for all gain and loss processes. The limits of these formulas for three- and four-level lasers are discussed. In Sec. 6, the main results of the tutorial are summarized. The tutorial also contains useful appendices on coherent and thermal states, the quantum fluctuations induced by gain and loss, the origin of the van der Pol equation and the integral of colored noise. It is written at a level that is suitable for last-year undergraduate students or first-year graduate students.

\section{Amplitude equations}

Our semi-classical model of laser evolution is based on the stochastic amplitude equation
\be d_t A = (\al - \bt)A/2 + R_a(t), \label{2.1} \ee
where $A$ is the complex wave amplitude, and $\al$ and $\bt$ are the complex gain and loss rates, respectively. The squared amplitude $|A|^2$ is the number of photons in the cavity. The parameter $\al$ represents stimulated emission, which depends on the number of upper-level (carrier) electrons, whereas $\bt = \bt_a + \bt_f + \bt_m$ represents stimulated absorption, which depends on the number of lower-level (valence) electrons, and facet and material loss, which do not depend on the electron numbers. We made the simplifying assumption that $\al$ and $\bt_a$ are constants, which requires the emission and absorption processes to have broad frequency bandwiths. The rate-equation model of a laser involves the electron and photon numbers, and the coefficients $\al_r$ and $\bt_r$, where the subscript $r$ denotes a real part. However, because gain and loss are causal processes, the Kramers--Kronig relation requires the amplitude gain and loss coefficients to have imaginary parts \cite{jac99}.

The (Langevin) source function $R_a(t)$ mimics the effects of quantum fluctuations (including spontaneous emission). It is a random function of time, with the properties
\be \<R_a(t)\> = 0, \ \ \<R_a(t)R_a(t')\> = 0, \ \ \<R_a^*(t)R_a(t')\> = \si_a\de(t - t'), \label{2.2} \ee
where $\<\ \>$ denotes an ensemble average and the source strength $\si_a = (\al_r + \bt_r)/2$. Equations (\ref{2.2}) are the defining properties of white noise. The first equation states that positive and negative, real and imaginary, impulses are equally likely. The second equation states that the real and imaginary impulses have equal strengths ($\si_a/2$), but are statistically independent. The third equation states that impulses at different times are independent.  Notice that the source strength depends on the sum of the gain and loss coefficients (not the difference). We assume that the (Wiener) increment $W(t) = \tint_0^t R_a(s)ds$ is a complex Gaussian random variable. %we assume that? 
Gaussian probability distributions are specified completely by their means and variances. It follows from Eqs. (\ref{2.2}) that
\be \<W(t)\> = 0, \ \ \<|W(t)|^2\> = \si_a t. \label{2.3} \ee
In App. B, it is shown that Eqs. (\ref{2.1}) and (\ref{2.2}) model accurately the amplitude (quadrature) fluctuations induced by gain and loss processes. 
%[NB: The formula for $\si_a$ only applies to a vacuum-state idler. Its generalization for a thermal-state idler should be stated later.]

Because the source functions vary randomly with time, so also does the driven amplitude, which is characterized by its moments in the time and frequency domains. The first time-domain moment $\<A(t)\>$, where $\<\ \>$ denotes an ensemble average, is called the mean and the second moment $\<A^*(t)A(t')\>$ is called the two-time correlation. Because noise is always present, the driven amplitude is not an integrable function of time. To avoid singularities in the theory, we use the finite-time Fourier transforms
\ba A(\om) &= &\int_0^T A(t) e^{i\om t} dt, \label{2.4} \\
A(t) &= &\int_{-\infty}^\infty A(\om) e^{-i\om t} d\om/2\pi. \label{2.5} \ea
In a previous tutorial \cite{mck21}, we discussed finite-time transforms and applied them to studies of simple harmonic oscillators driven by noise. Many of the mathematical results and physical insights described therein are also relevant to lasers.
In this tutorial, $A(t)$ is dimensionless, so $A(\om)$ has units of time. However, in some applications $|A(t)|^2$ has units of power, in which case $|A(\om)|^2$ has units of energy multiplied by time (divided by frequency). For this reason, $|A(\om)|^2$ is called the spectral energy density. %and $|A(\om)|^2/T$ is called the spectral power density.
We will use the descriptive terms power and energy whenever it is helpful to do so.
It follows from Eq. (\ref{2.4}) that the second frequency-domain moment
\be \<|A(\om)|^2\> = \int_0^T \int_0^T \<A^*(t)A(t')\> e^{i\om(t' - t)} dt dt'. \label{2.6} \ee
Thus, if the temporal correlation is known, the frequency spectrum can be calculated.

\section{Below-threshold laser}

First, consider a well-below-threshold laser ($\al_r \ll \bt_r$). In this regime, the number of photons is too small to affect the numbers of upper- and lower-level electrons, which are determined by the pump power (electron current) and the relevant electron-decay processes. Let $A(t) = B(t) e^{i(\al_i - \bt_i)t/2}$, where the subscript $i$ denotes an imaginary part. Then the transformed amplitude satisfies the equation
\be d_t B = -\nu_r B + R_b(t), \label{3.1} \ee
where $\nu_r = (\bt_r - \al_r)/2$ is the net-damping rate and $R_b(t) = R_a(t) e^{i(\bt_i - \al_i)t/2}$ is the transformed source function. It is easy to verify that $R_b$ also has properties (\ref{2.2}), so the phase factor can be omitted (and $\si_b = \si_a$). Equation (\ref{3.1}) has the same form as the equation for a strongly-damped oscillator driven by noise, which was studied in Sec. 3 of \cite{mck21}.

The solution of Eq. (\ref{3.1}) is
\be B(t) = B(0)e^{-\nu t} + \tint_0^t R_b(s)e^{-\nu(t - s)} ds, \label{3.2} \ee
where the subscript $r$ was omitted. Typically, the growth of the amplitude is initiated by noise, in which case $B(0) = 0$. One could mimic the fluctuations associated with an initial vacuum state by setting $\<B(0)\> = 0$ and $\<|B(0)|^2\> = 1/2$ (App. A), but this contribution to the amplitude is damped and does not affect the stochastic steady state, so we will omit it. The remaining (driven) part of solution (\ref{3.2}) has the properties
\ba \<B(t)\> &= &0, \label{3.3} \\
\<|B(t)|^2\> &= &\int_0^t \int_0^t \<R_b^*(s)R_b(s')\> e^{-\nu(2t - s - s')} dsds' \nonumber \\
&= &\int_0^t \si_b e^{-2\nu(t - s)} ds \nonumber \\
&= &\si_b(1 - e^{-2\nu t})/2\nu. \label{3.4} \ea
Equation (\ref{3.3}) shows that the amplitude mean is zero, whereas Eq. (\ref{3.4}) shows that, for times longer than the damping time $1/\nu$, the amplitude variance (number mean) tends to a steady-state value, which is proportional to the source strength and inversely proportional to the damping rate. For $t' > t$, the temporal correlation
\ba \<B^*(t)B(t')\> &= &\int_0^t \int_0^{t'} \<R_b^*(s)R_b(s')\> e^{-\nu(t + t' - s - s')} ds'ds \nonumber \\
&= &\int_0^t \si_b e^{-\nu(t + t' - 2s)} ds \nonumber \\
&= &\si_b [e^{-\nu(t' - t)} - e^{-\nu(t + t')}]/2\nu. \label{3.5} \ea
%
%(NB: I could omit the first two lines of this equation.)
Because correlation (\ref{3.5}) is real, it is a symmetric function of $t$ and $t'$. Hence, for $t > t'$, the first exponent in Eq. (\ref{3.5}) must be $-\nu(t - t')$. For times that are longer then the damping time,
\be \<B^*(t)B(t')\> \approx \si_b e^{-\nu|t - t'|}/2\nu. \label{3.6} \ee
Equation (\ref{3.6}) is a defining property of colored noise. It shows that that correlation (coherence) time of the amplitude is the damping time. A system whose first two moments depend on the time difference, but not the time origin, is in a (weakly) stationary state. For such a system, $\<B(t)B(t')\> = \<B(t)B(t + \ta)\> = \<B(0)B(\ta)\>$, where $\ta = t' - t$ is the time difference. 

One obtains a formula for the spectrum by combining Eqs. (\ref{2.6}) and (\ref{3.6}). The relevant integrals were done in Sec. 2 of \cite{mck21} and the result is
\be \<|B(\om)|^2\> \approx \frac{\si_b}{\nu} \Biggl[\frac{\nu T}{\nu^2 + \om^2} - \frac{\nu^2 - \om^2}{(\nu^2 + \om^2)^2}\Biggr]. \label{3.7} \ee
For long measurement times ($\nu T \gg 1$), the spectrum is asymptotic to the Lorentzian
\be \<|B(\om)|^2\> = \si_bT/(\nu^2 + \om^2), \label{3.8} \ee
which has a FWHM of $2\nu$.
Although the spectral energy density increases in proportion to $T$ (which is natural for a sytem that is driven continuously by noise), the spectral power density $ \<|B(\om)|^2\>/T$ is constant. (One can verify the formula for the power density by extending the limits of the $t - t'$ integral to $\pm \infty$ and normalizing the result.) %and applying the Wiener--Khinchine theorem.)

The original amplitude $A$ differs from the transformed amplitude by the time-dependent phase factor $e^{-i\nu_it}$. It follows from this observation, and Eqs. (\ref{3.6}) and (\ref{3.8}), that the original correlation and spectrum are
\ba \<A^*(0)A(\ta)\> &= &(\si_a/2\nu_r) e^{-\nu_r|\ta| - i\nu_i\ta}, \label{3.9} \\
\<|A(\om)|^2\> &= &\si_aT/[\nu_r^2 + (\om - \nu_i)^2], \label{3.10} \ea
respectively, where the subscript $r$ was restored. The original correlation is phase shifted and the original spectrum is frequency shifted, but the correlation time and spectrum width still depend on $\nu_r$ only.

%(NB: Nowhere else in this short tutorial do I mention number fluctuations, so this section is out of place. However, it is short and interesting, and does not fit in anything else I plan to write, so I am inclined to keep it, to complete the below-threshold story.)
Solution (\ref{3.2}) shows that the amplitude is the sum of impulses, which have Gaussian statistics, multiplied by deterministic damping factors. %(Green functions).
Hence, the amplitude also has Gaussian statistics and its probability distribution is specified completely by Eqs. (\ref{3.3}) and (\ref{3.4}). Let $x$ and $y$ denote the real and imaginary parts of $B$, respectively, and let $\si = \si_b/4\nu$. Then the probability distribution function 
\be P(x,y) = \exp[-(x^2 + y^2)/2\si]/(2\pi\si). \label{3.11} \ee
Now let $F(x,y)$ be an arbitrary function of its arguments. Then the average
\be \<F(x,y)\> = \int\int F(x,y)P(x,y) dxdy. \label{3.12} \ee
It is easy to verfy that $\<x^2\> = \<y^2\> = \si$, so the number mean $\<x^2 + y^2\> = 2\si = \si_b/2\nu$.

It is instructive to consider the distribution in a different coordinate system. Let $u$ and $v$ be arbitrary functions of $x$ and $y$. Then
\be P(u,v)dudv = P(u,v)J(u,v|x,y)dx dy = P(x,y)dx dy, \label{3.13} \ee
where $J$ is the Jacobian of the transformation. Hence,
\be P(u,v) = P(x,y)/J(u,v|x,y), \label{3.14} \ee
where the functions on the right side are written in terms of $u$ and $v$. For polar coordinates, $u = n = x^2 + y^2$, $v = \ph = \tan^{-1}(y/x)$ and $J = 2$. It follows from Eqs. (\ref{3.11}) and (\ref{3.14}) that
\be P(n,\ph) = \exp(-n/2\si)/(4\pi\si). \label{3.15} \ee
Distribution (\ref{3.15}) is independent of phase, so the number distribution
\be P(n) = \exp(-n/\si_n)/\si_n, \label{3.16} \ee
where $\si_n = 2\si$. A Gaussian amplitude distribution is equivalent to an exponential number distribution and a uniform phase distribution. It follows from Eq. (\ref{3.16}) that the number mean $\<n\> = \si_n$ and the number variance $\<\de n^2\> = \<n^2\> - \<n\>^2 = \si_n^2$. A below-threshold laser emits light for which the number deviation is as large as the number mean. This behavior is not useful. It contrasts markedly with the behavior of an above-threshold laser, for which the deviation is much smaller than the mean.

\section{Above-threshold laser}

As one increases the pump power, the number of upper-level electrons increases. At some point (threshold), the gain $\re(\al_e)$ equals the total loss $\re(\bt_a) + \bt_f + \bt_m$. A further increase in pump power causes photons to be generated, first by spontaneous emission and then by stimulated emission. Because the generation of a photon coincides with the lowering of an electron, the number of upper-level electrons decreases. There follows a sequence of (diminishing) relaxation oscillations in the electron and photon numbers. After some time, the laser reaches a self-consistent equilibrium, in which
\be \re[\al_e(|A_0|^2)] = \re[\bt_a(|A_0|^2)] + \bt_f + \bt_m, \label{4.1} \ee
where $A_0$ is the equilibrium amplitude. Condition (\ref{4.1}) is called gain clamping. It constrains the real parts of the gain and loss coefficients, but does not constrain the imaginary parts. In equilibrium, photons leave the cavity at the rate $\bt_f|A_0|^2$. Excess pump power is converted to laser power. Although the gain and loss processes are in balance, they still produce noise, so the output amplitude fluctuates. These fluctuations are the cause of laser linewidth.

In the vicinity of equilibrium, the net loss
\be \nu(|A|^2) \approx 0 + i\nu_{0i} + \nu_0'(|A|^2 - |A_0|^2), \label{4.2} \ee
where $\nu_{0i} = \nu_i(|A_0|^2)$ and $\nu_0' = d\nu/d|A|^2$, evaluated at $A_0$. Equation (\ref{4.2}) is based on the assumption that the active medium responds instantaneously to changes in photon number. By combining it with Eq. (\ref{2.1}), one finds that
\be d_t A = -i\nu_{0i}A - \nu_0'(|A|^2 - |A_0|^2)A + R_a(t), \label{4.3} \ee
where $R_a$ has properties (\ref{2.2}), which depend on the equilibrium gain and loss rates. The first term on the right side of Eq. (\ref{4.3}) produces a frequency shift, which is of secondary importance. One can eliminate it by making the transformation $A(t) = B(t)e^{-i\nu_{0i}t}$, in which case
\be d_t B = \nu_0'(|B_0|^2 - |B|^2)B + R_b(t), \label{4.4} \ee
where the source function $R_b(t) = R_a(t)e^{i\nu_{0i}t}$ also has properties (\ref{2.2}). In the optics literature, this (first-order) equation is called the van der Pol (vdP) equation. Its relation to the (second-order) electrical--mechanical vdP equation is described in App. C.

In equilibrium and in the absence of noise, $B(t) = B_0$ is a constant (which one can choose to be real and positive). To study small perturbations of this equilibrium in the presence of noise, let $B(t) = B_0 + B_1(t)$, where $|B_1| \ll B_0$. Then, by making this substitution in Eq. (\ref{4.4}), and splitting the resulting equation into real and imaginary parts, one obtains the linearized equations
\ba d_t B_{1r} &= &-\mu_r B_{1r} + R_r(t), \label{4.5} \\
d_t B_{1i} &= &-\mu_i B_{1r} + R_i(t), \label{4.6} \ea
where $\mu_r = 2\nu_{0r}'|B_0|^2$ and $\mu_i = 2\nu_{0i}'|B_0|^2$. The source terms $R_r(t)$ and $R_i(t)$ are real and have the properties
\be \<R_j(t)\> = 0, \ \ \<R_j(t)R_k(t')\> = \si_j\de_{jk}\de(t - t'), \label{4.7} \ee
where $\si_r = \si_i = \si_b/2$. Notice that the real amplitude is damped and the imaginary amplitude is coupled to the real amplitude by the imaginary part of the net-loss coefficient. The solutions of Eqs. (\ref{4.5}) and (\ref{4.6}) are
\ba B_{1r}(t) &= &\int_0^t R_r(s) e^{-\mu_r(t - s)} ds, \label{4.8} \\
B_{1i}(t) &= &\int_0^t [\mu_i B_{1r}(s) + R_i(s)] ds. \label{4.9} \ea

First consider the real amplitude. Solution (\ref{4.8}) has the same form as solution (\ref{3.5}), so the mean $\<B_{1r}(t)\> = 0$ and, for times longer than the damping time $1/\mu_r$, the temporal correlation
\be \<B_{1r}(t)B_{1r}(t')\> = (\si_r/2\mu_r) e^{-\mu_r|t - t'|}. \label{4.10} \ee
Although the below- and above-threshold (real) correlations are similar, their magnitudes and correlation times differ. [There is no simple relation between $\nu_r = (\bt_r -\al_r)/2$ and $\mu_r = 2\nu_{0r}'|B_0|^2$.] Once again, the real amplitude attains a (weakly) stationary state and has the properties of colored noise. Its deviation is much smaller than the equilibrium amplitude.

Second, consider the imaginary amplitude, which is the integral of a combination of white noise ($R_i$) and colored noise ($\mu_iB_{1r}$). The mean amplitude $\<B_{1i}(t)\> = 0$. The white-noise contribution to the correlation is
\ba \<B_{1i}(t)B_{1i}(t')\>_w &= &\int_0^t \int_0^{t'} \<R_i(s)R_i(s')\> ds'ds \nonumber \\
&= &\int_0^{\min(t,t')} \si_i ds \nonumber \\
&= &\si_i \min(t,t'). \label{4.11} \ea
This contribution grows without bound, because the imaginary amplitude is not damped. It never becomes a function of the time difference, so it never attains a stationary state. However, the noise source in Eq. (\ref{4.9}) is stationary, so the contribution has stationary increments.
%Correlation (\ref{4.11}) is proportional to the minimal time, but does not depend on the time origin.

The colored-noise contribution is determined exactly in App. D. One can estimate it by observing that right side of Eq. (\ref{4.10}) can be rewritten as $(\si_r/\mu_r^2)\de_e(\ta)$, where, for times that are longer than the correlation time, $\de_e(\ta) = (\mu_r/2) e^{-\mu_r|\ta|}$ is an effective $\de$-function. (It has a narrow peak and an integral of 1.) For such times, it follows from Eqs. (\ref{4.9}) and (\ref{4.11}) that the colored-noise contribution is
\be \<B_{1i}(t)B_{1i}(t')\>_c \approx (\si_r\mu_i^2/\mu_r^2) \min(t,t'). \label{4.12} \ee
This contribution is also nonstationary.
By combining Eqs. (\ref{4.11}) and (\ref{4.12}), and using the fact that $\si_r = \si_i$, one obtains the correlation
\be \<B_{1i}(t)B_{1i}(t')\> = \si_i(1 + \ep^2)\min(t,t'), \label{4.13} \ee
where $\ep = \mu_i/\mu_r$ is the enhancement factor. Notice that $\mu_i/\mu_r = \nu_{0i}'/\nu_{0r}'$. (The factors of $|B_0|^2$ cancel.) In App. D, it is shown that the error in formula (\ref{4.13}) is of relative order $1/\mu_rt$, which is small by assumption.

One obtains formulas for the spectra of the real and imaginary amplitudes by combining Eq. (\ref{2.6}) with Eqs. (\ref{4.10}) and (\ref{4.13}), respectively. The first calculation was described briefly in the previous subsection. It follows from Eq. (\ref{3.8}) that the real spectrum
\be \<|B_{1r}(\om)|^2\> = \si_r T/(\mu_r^2 + \om^2). \label{4.14} \ee
Although the below- and above-threshold (real) spectra are both Lorentzian, their strengths and widths differ significantly.
The second calculation was done in Sec. 5 of \cite{mck21}. The result is the imaginary spectrum
\be \<|B_{1i}(\om)|^2\> = 2\si_iT(1 + \ep^2)[1 - \sinc(\om T)]/\om^2. \label{4.15} \ee
For very low frequencies ($\om \ll 1/T$), the energy density is approximately $\si_i(1 + \ep^2)T^3/3$. The associated power density is proportional to $T^2$, which tends to infinity as $T$ tends to infinity. It is for this reason that we use finite measurement times. For typical frequencies ($\om \gg 1/T$), the sinc term is negligible and the spectrum is proportional to $1/\om^2$. This frequency dependence is a characteristic of Brownian noise.

Previous researchers rewrote the vdP equation (or similar equations) in terms of the number $N = |B|^2$ and phase $\ph = \tan^{-1}(B_i/B_r)$. However, this change of variables is nonlinear. It converts a pair of linear equations with additive noise to a pair of nonlinear equations with multiplicative noise. For such equations, one must follow the rules of stochastic calculus carefully \cite{gar85}. In the context of this tutorial, we thought that stochastic calculus would be an unnecessary distraction, so we chose to continue working with the real and imaginary amplitudes. Previous researchers derived number spectra that are Lorentzian \cite{vah83a,vah83b,spa83} and a phase spectrum that decreases as $1/\om^2$ \cite{hen83}. Equations (\ref{4.14}) and (\ref{4.15}) are consistent with these results.

As long as the amplitude perturbations remain smaller than the equilibrium amplitude, the probability distribution function is a two-dimensional Gaussian. The real (in-phase) part of the amplitude is centered on $B_0$ and has the variance $\si_r/2\mu_r$, which is constant, whereas the imaginary (out-of-phase) part is centered on 0 and has the variance $\si_i(1 + \ep^2)t$, which increases linearly with time. This distribution function resembles that of an amplitude-squeezed (phase-stretched) state, which is illustrated in \cite{lou00}. 

The original amplitude $A$ is phase-shifted version of the transformed amplitude $B$, so the original correlations are phase-shifted versions of correlations (\ref{4.10}) and (\ref{4.13}), and the original spectra are frequency-shifted versions of spectra (\ref{4.14}) and (\ref{4.15}). The center of the original amplitude distribution rotates.

\section{Laser linewidths}

The laser linewidth $\de\om$ is defined to be the FWHM of the amplitude spectrum. In Sec. 3, it was shown that in the well-below-threshold regime, $\de\om = 2\nu$ is the (number) damping rate $\bt - \al$ [Eq. (\ref{3.8})], where $\al \ll \bt$. It was also shown that the photon number $N = \si/2\nu$, where the source strength $\si = (\al + \bt)/2$ [Eq. (\ref{3.4})]. Thus, the linewidth equals the damping rate, and the number is proportional to the source strength and inversely proportional to the damping rate. As the pump power increases, $\al$ increases and $\bt_a$ decreases, so $\nu$ decreases. This behavior is called line narrowing. As $\nu$ decreases, $N$ increases concomitantly. At some point, the approximation that $\al$ and $\bt_a$ are independent of $N$ ceases to be accurate, but by this point the trends in linewidth and number are clear. For us, it is natural to characterize a below-threshold laser in terms of the damping rate and source strength. However, previous authors characterized it in terms of the output power. By using the facts that $2\nu = \si/N$ and $P = N\hb\om\bt_f$, one obtains the below-threshold linewidth formula
\be \de\om = \hb\om\bt_f(\al + \bt)/2P. \label{5.1} \ee
Equation (\ref{5.1}) includes the effects of emission, absorption, and facet and material loss. The linewidth narrows as the power increases. Well below threshold, $\al \ll \bt$. For a four-level laser (without absorption or material loss),
\be \de\om_4 = \hb\om\bt_f^2/2P. \label{5.2} \ee
For a three-level laser (with absorption, but without material loss),
\be \de\om_3 = \hb\om\bt_f(\bt_a + \bt_f)/2P. \label{5.3} \ee
Thus, three-level lasers have broader linewidths than four-level lasers (with the same $\bt_f$).
%(NB: I think the GZT and ST formulas are BT formulas.)

In the above-threshold regime, the complex amplitude $B_0 + B_{1r}(t) + iB_{1i}(t)$ is the sum of constant and fluctuating terms. In Sec. 4, it was shown that the real amplitude perturbation is damped and bounded, whereas the imaginary perturbation is undamped and unbounded. Under these circumstances, it is reasonable to model complex amplitude fluctuations as phase fluctuations, where the phase $\ph(t) \approx B_{1i}(t)/B_0$. It follows from Eq. (\ref{4.13}) that the phase mean $\<\ph(t)\> = 0$ and the phase variance
\be \<\ph^2(t)\> = \si_i(1 + \ep^2)t/|B_0|^2. \label{5.4} \ee

The amplitude correlation $\<B^*(0)B(t)\> = |B_0|^2 \<e^{i\ph(t)}\>$. The phase is the time integral of impulses that have Gaussian probability distributions [Eq. (\ref{4.9})], so at any time, the phase also has a Gaussian distribution. For such a distribution,
\ba \<e^{ix}\> &= &\int_{-\infty}^\infty e^{ix} e^{-x^2/2\si_x} dx/(2\pi\si_x)^{1/2} \nonumber \\
&= &\int_{-\infty}^\infty e^{-\si_x/2} e^{-(x - i\si_x)^2/2\si_x} dx/(2\pi\si_x)^{1/2} \nonumber \\
&= &e^{-\si_x/2}, \label{5.5} \ea
where $\si_x$ is the variance.
By combining Eqs. (\ref{5.4}) and (\ref{5.5}), one obtains the correlation
\be \<B^*(0)B(t)\> = |B_0|^2 \exp[-\si_i(1 + \ep^2)t/2|B_0|^2]. \label{5.6} \ee
This correlation decreases exponentially in time, at the rate  $\ga = \si_i(1 + \ep^2)/2|B_0|^2$. It follows from Eqs. (\ref{3.6}) and (\ref{3.8}) that the amplitude spectrum
\be \<|B(\om)|^2\> = 2|B_0|^2\ga T/(\ga^2 + \om^2).  \label{5.7} \ee
Thus, the FWHM of the spectrum is $\si(1 + \ep^2)/2N$, where $\si = 2\si_i$ and $N = |B_0|^2$. 
Although the original amplitude $A$ is a phase-shifted version of the transformed amplitude $B$, both amplitudes have the same linewidth, as explained after Eqs. (\ref{3.10}) and (\ref{4.15}).

By converting from photon number to output power (as described above), one obtains the above-threshold linewidth formula
\be \de\om = \hb\om\bt_f(\al + \bt)(1 + \ep^2)/4P. \label{5.8} \ee
Equation (\ref{5.2}) also includes the effects of emission, absorption, and facet and material loss. In equilibrium, $\al = \bt$, so addition (emission) and subtraction (absorption, and facet and material loss) contribute equally to the linewidth.
%(NB: In the BT regime, only loss contributes, so this equality of contributions produces a factor of 2.)

The above-threshold formula (\ref{5.8}) differs from the below-threshold formula (\ref{5.1}) in three significant ways. First (as mentioned above), gain and loss contribute equally to the linewidth. Second, the linewidth is proportional to $1 + \ep^2$, where $\ep = \nu_{0i}'/\nu_{0r}' = -\ch_r^{(3)}/\ch_\imath^{(3)}$ is the enhancement factor. For semiconductor lasers, this factor is signficant ($|\ep| \approx 5$). In the remainder of this section, the enhancement factor will be omitted (because it appears in all the formulas and previous researchers agree upon it). Third, the denominator in Eq. (\ref{5.8}) is twice as large as the denominator in Eq. (\ref{5.1}). The below- and above-threshold calculations are different, so there is no simple relation between them (in the context of a phenomenological vdP equation). However, the fact that only imaginary amplitude perturbations affect the above-threshold linewidth does contribute a factor of 1/2 to its final value.

For a four-level laser (without absorption or material loss), the linewidth
\be \de\om_4 = \hb\om\bt_f^2/2P. \label{5.9} \ee
Emission and facet loss contribute equally to the linewidth, so $(\al + \bt_f)/4 = \bt_f/2$. Hence, the above-threshold formula is identical to the well-below-threshold formula.
For a three-level laser (with absorption, but without material loss), $\de\om = \hb\om\bt_f(\bt_a + \bt_f)/2P$. Once again, three-level lasers have broader linewidths than comparable four-level lasers, and the above-threshold formula is identical to the well-below-threshold formula. It follows from the equilibrium condition that $\bt_a + \bt_f = \al$ and $\al/\bt_f = \al/(\al - \bt_a)$. By using these facts, one can rewrite the linewidth formula as
\be \de\om_3 = [\al/(\al - \bt_a)](\hb\om\bt_f^2/2P). \label{5.10} \ee
By using the facts that $\al = gN_2$ and $\bt_a = gN_1$, where $g$ is the transition coefficient, and $N_1$ and $N_2$ are the lower- and upper-level electron numbers, respectively, one can rewrite the first factor in Eq. (\ref{5.10}) as $N_2/(N_2 - N_1)$. This factor is called the spontaneous noise factor ($n_{{\rm sp}}$).

Without the enhancement factor, formula (\ref{5.9}) is equivalent to the short HLS formula, which also includes the effects of gain and loss fluctuations. With the enhancement factor, it has the same form as the Henry formula. However, Henry chose to include only gain fluctuations. In order to reproduce the spontaneous contribution to the photon-generation rate $\al(N + 1)$, he used the semi-classical strength $\si_\al = \al$ (which is twice our strength). This choice leads to the incremental quadrature variance $\<|\de A|^2 + |\de A|^2\>/2$, which is proportional to $2\si_\al/2 = \al$. When one does the same calculation quantum mechanically (App. B), one obtains the incremental variance $\<\de a^\d\de a + \de a\de a^\d\>/2$, which is proportional to $\al/2$ (because only the second term contributes). So Henry overestimated the quadrature gain fluctuations, which compensated for his omission of loss fluctuations.

The preceding linewidth formulas are all based on the quantum fluctuations associated with gain ($\si_\al = \al/2$) and loss ($\si_\bt = \bt/2$). One should also consider thermal fluctuations. The properties of a thermal quantum state are described in Apps. A and B. One can mimic the effects of thermal fluctuations by adding  to Eq. (\ref{2.1}) a source of strength $\bt\<n_r\>$, where $\<n_r\> = 1/(e^{\hb\om/kT} - 1)$ is the expected number of reservoir photons. In the absence of gain, this source allows the cavity mode to reach equilibrium with the reservoir, as required by the fluctuation--dissipation theorem. Thus, one can quantify the effects of thermal fluctuations on the linewidth by making the substitution $\bt/2 \rightarrow \bt(1/2 + \<n_r\>) = \bt(1 + 2\<n_r\>)/2$ in Eq. (\ref{5.8}). By including the thermal source associated with facet loss, which is the loss of a passive cavity (without the active medium), one obtains the generalized linewidth formula
\be \de\om = \frac{\hb\om\bt_f^2}{4P} \Biggl(\frac{\al + \bt_a}{\al - \bt_a} + 1 + 2\<n_r\>\Biggr). \label{5.11} \ee
Formula (\ref{5.11}) is equivalent to the long HLS formula. However, one might argue that the thermal strength should be proportional to $\bt = \bt_a + \bt_f + \bt_m$, which is the total cavity loss (with the active medium, but without the pump).

For light waves, $\<n\> \approx e^{-\hb\om/kT} \ll 1$, in which case formula (\ref{5.11}) reduces to formulas (\ref{5.9}) and (\ref{5.10}) in the appropriate limits. In contrast, for microwaves, $\<n_r\> \approx kT/\hb\om \gg 1$, in which case formula (\ref{5.11}) reduces to
\be \de\om = kT\bt_f^2/2P. \label{5.12} \ee
Formula (\ref{5.12}) is similar to formula (\ref{1.1}), with the emission bandwidth $\De\om$ replaced by the cavity bandwidth $\bt_f$.

\section{Summary}

In this tutorial, the physical origins and mathematical analyses of laser linewidths were reviewed. Our semi-classical model is based on an equation for the light-mode amplitude [Eq. (\ref{2.1})] that includes random source terms, one term for each process that affects the amplitude (stimulated and spontaneous emission, stimulated absorption, and facet and material loss). Although the source terms are classical, their assigned strengths [Eqs. (\ref{2.2})] are consistent with the laws of quantum optics [Eqs. (\ref{b28}) and (\ref{b31})].
Analyses of this equation show that the below- and above-threshold laser linewidths [Eqs. (\ref{5.1}) and (\ref{5.8})] are proportional to the sum of the (positive) source strengths for all gain and loss processes, and inversely proportional to the output power.
Three-level and semiconductor lasers have broader linewidths than comparable four-level lasers, because stimulated absorption and the stimulated emission that compensates it both contribute to the linewidths.
In the below-threshold regime, the complex amplitude spectrum is Lorentzian [Eq. (\ref{3.8})]. In the above-threshold regime, the real amplitude (power) spectrum is Lorentzian [Eq. (\ref{4.14})], whereas the imaginary amplitude (phase) spectrum is Brownian [Eq. (\ref{4.15})]. Our above-threshold linewidth formula [Eq. (\ref{5.8})] is consistent with the Haken--Lax--Scully formula \cite{hak66,lax66,scu66}, and includes the linewidth-enhancement factor of Haug \cite{hau67} and Henry \cite{hen82}.

\section*{Appendix A: Coherent and thermal states}

In this appendix, some properties of coherent and thermal states are derived and used to gauge the accuracy of the semi-classical model of amplitude fluctuations. Let $a$ be a mode-amplitude operator, which satisfies the boson commutation relations
\be [a, a] = 0, \ \ [a, a^\d] = 1, \label{a1} \ee
where $[x, y] = xy - yx$ is a commutator and $\d$ denotes a hermitian conjugate. The quadrature operator $q(\ph) = (a^\d e^{i\ph} + a e^{-i\ph})/2^{1/2}$, where $\ph$ is the phase of the local oscillator, and the number operator $n = a^\d a$.

A coherent state is defined by the eigenvalue equation
\be a|\al\> = \al|\al\>, \label{a2} \ee
where $|\al\>$ is the state vector and $\al$ is a complex number. It follows from Eq. (\ref{a2}) that the expectation value $\<\al|a|\al\> = \<a\> = \al$, so $\al$ is the amplitude mean.
The quadrature mean
\be \<q(\ph)\> = (\al^*e^{i\ph} + \al e^{-i\ph})/2^{1/2}. \label{a3} \ee
By combining Eqs. (\ref{a1}) and (\ref{a2}), one finds that the second quadrature moment
\ba \<q^2(\ph)\> &= &\<(a^\d)^2e^{i2\ph} + a^\d a + aa^\d + a^2e^{-i2\ph}\>/2 \nonumber \\
&= &[(\al^*)^2e^{i2\ph} + |\al|^2 + (|\al|^2 + 1) + \al^2e^{-i2\ph}]/2. \label{a4} \ea
The quadrature deviation $\de q = q - \<q\>$, so the quadrature variance $\<\de q^2\> = \<q^2\> - \<q\>^2$. By combining Eqs. (\ref{a3}) and (\ref{a4}), one finds that the quadrature variance
\be \<\de q^2(\ph)\> = 1/2. \label{a5} \ee
Notice that the variance is phase independent. For a vacuum state ($\al = 0$), the quadrature mean is zero, but the quadrature variance is nonzero. These quadrature fluctuations are called vacuum fluctuations.
It follows from the second term on the right side of Eq. (\ref{a4}) that the number mean $\<n\> = |\al|^2$. In \cite{lou00}, the number variance is shown to equal the number mean.

A coherent state is pure, so its density matrix $\rh = |\al\>\<\al|$. Let $o$ be an arbitrary operator. Then its expected value is the trace $\tr(\rh o)$. By using the identity $\tr(|x\>\<y|) = \<y|x\>$, one finds that $\tr(\rh o) = \<\al|o|\al\>$. For a pure state, the definitions of expectation value are equivalent. Now let $\{|\psi_n\>\}$ be a complete set of state vectors, which are normalized, but are not necessarily orthogonal. Then a mixed state is specified by the density matrix $\rh = \tsum_n p_n |\ps_n\>\<\ps_n|$, where the total probability $\tsum_n p_n = 1$. By using the trace identity, one finds that $\tr(\rh o) = \tsum_n p_n \<\ps_n|o|\ps_n\>$. The expectation value is the weighted sum of the expectation values associated with the basis states.

A thermal state is defined by the density matrix
\be \rh = \tsum_n p_n |n\>\<n|, \label{a11} \ee
where $|n\>$ is a number state. The probability $p_n = (1 - x)x^n$, where $x = e^{-\hb\om/kT}$ is the Boltzmann factor. This probability distribution is called the Planck %(or Bose--Einstein)
distribution. By using the fact that $\tsum_n x^n = 1/(1 - x)$, one can verify that the total probability is 1. The number mean
\ba \<n\> &= &(1 - x) \tsum_n nx^n \ = \ (1 - x) xd_x \tsum_n x^n \nonumber \\
&= &(1 - x) xd_x [1/(1 - x)] \ = \ x/(1 - x). \label{a12} \ea
By inverting Eq. (\ref{a12}), one finds that $x = \<n\>/(1 + \<n\>)$. The properties of a thermal state are determined completely by its number mean. By writing the Boltzmann factor explicitly, one finds that
\be \<n\> = 1/(e^{\hb\om/kT} - 1). \label{a15} \ee
For low frequencies (microwaves), $\<n\> \approx kT/\hb\om \gg 1$, whereas for high frequencies (light waves) $\<n\> \approx e^{-\hb\om/kT} \ll 1$.

Density matrix (\ref{a11}) is diagonal in the number-state basis. Hence, operators that change diagonal elements to non-diagonal ones (which do not appear in traces) have expectation values of zero. It follows from this observation that the quadrature mean
\be \<q(\ph)\> = 0. \label{a13} \ee
In this case, the quadrature variance equals the second quadrature moment. It follows from the first line of Eq. (\ref{a4}) that
\be \<\de q^2(\ph)\> = \<n\> + 1/2. \label{a14} \ee
Notice that the quadrature mean and variance are both phase independent. The quadrature variance of a thermal state exceeds that of a coherent state. In \cite{lou00}, the number variance of a thermal state is shown to be $\<n\>^2 + \<n\>$, which also exceeds that of a coherent state.

In the semi-classical model of amplitude fluctuations, one replaces the mode operator $a$ by the amplitude $A + \de A$, where $A = \<a\>$ is the mean amplitude and $\de A$ is a Gaussian random variable with the properties
\be \<\de A\> = 0, \ \ \<\de A^2\> = 0, \ \ \<|\de A|^2\> = \si, \label{a21} \ee
where $\<\ \>$ denotes an ensemble average. The variance $\si$ remains to be determined. The semi-classical quadrature is defined in the same way as the quantum one. It follows from the first of Eqs. (\ref{a21}) that the quadrature mean
\be \<Q(\ph)\> = (A^* e^{i\ph} + A e^{-i\ph})/2^{1/2}. \label{a22} \ee
Equation (\ref{a22}) is equivalent to Eq. (\ref{a3}), which applies to a coherent state. The quadrature deviation $\de Q = (\de A^* e^{i\ph} + \de A e^{-i\ph})/2^{1/2}$, from which it follows that the quadrature variance
\be \<\de Q^2(\ph)\> = \si. \label{a23} \ee
One can reconcile Eq. (\ref{a23}) with Eq. (\ref{a5}) by setting $\si = 1/2$.
One can model the properties of a thermal state by setting $A = 0$ and $\si = \<n\> + 1/2$. By doing so, one reproduces the quadrature mean (\ref{a13}) and variance (\ref{a14}). Thus, the semi-classical model predicts the quadrature properties of coherent and thermal states accurately. (Further analysis shows that it overestimates the number means by 1/2 and the number variances by 1/4.)

In Sec. 2, we assumed that the amplitude (quadrature) fluctuations have Gaussian statistics. For coherent and thermal states, the quadrature distributions are Gaussian, with variances of 1/2 and $\<n\> + 1/2$, respectively \cite{gar08,mck12}, so our assumption was valid.

\section*{Appendix B: Quantum fluctuations}

Consider the evolution of a light mode in the presence of gain and loss. These processes involve auxiliary modes, which interact with the light mode. One can avoid modeling the interactions explicitly by using the stochastic amplitude-operator equation
\be d_t a = (\al - \bt)a/2 + s_\al^\d(t) + s_\bt(t), \label{b1}\ee
where $\al$ and $\bt$ are the gain and loss coefficients (rates), respectively. The noise operators $s_\al$ and $s_\bt$ are random operator functions of time, which model the effects of auxiliary-mode fluctuations \cite{mck20a,cav82}. Like their classical counterparts, they are independent, and their first and second squared moments are
\be \<s(t)\> = 0, \ \ \<s(t)s(t')\> = 0, \label{b2} \ee
where $\<\ \>$ denotes an ensemble average. Their mixed second moments are
\be \<s^\d(t)s(t')\> = \si_{\rm n}\de(t - t'), \ \ \<s(t)s^\d(t')\> = \si_{\rm a}\de(t - t'), \label{b3} \ee
where $\si_{\rm n}$ and $\si_{\rm a}$ are the normal and anti-normal strength coefficients, respectively.
The values of these coefficients are fixed by the requirement that the amplitude operators satisfy the commutation relations for all times. (Interactions with auxiliary modes are unitary.) Notice that Eq. (\ref{b1}) involves $s_\al^\d$ and $s_\bt$. The presence (absence) of the $\d$ accounts for the differences between the noise properties of gain and loss.

Let $\ta$ be a short time interval. Then the differential equation (\ref{b1}) is equivalent to the difference equation
\be a_\ta = a_0[1 + (\al - \bt)\ta/2] + \tint_0^\ta s_\al^\d(t) dt + \tint_0^\ta s_\bt(t) dt, \label{b4} \ee
where $a_\ta = a(\ta)$ \cite{gar85,mck20a}. The first term on the right side of Eq. (\ref{b4}) has contributions of order 1 and $\ta$. It follows from the second of Eqs. (\ref{2.3}) that the second and third terms are of order $\ta^{1/2}$.

Suppose, temporarily, that $\bt = 0$. Then, by combining Eq. (\ref{b4}) with its conjugate and retaining the squares of the noise terms, which are of order $\ta$, one finds that
\ba a_\ta^2 &= &a_0^2 (1 + \al\ta) + 2 a_0 \tint s_\al^\d + (\tint s_\al^\d)^2, \label{b5} \\
a_\ta^\d a_\ta &= &a_0^\d a_0 (1 + \al\ta) + a_0^\d \tint s_\al^\d + a_0 \tint s_\al + \tint s_\al \tint s_\al^\d, \label{b6} \\
a_\ta a_\ta^\d &= &a_0a_0^\d (1 + \al\ta) + a_0^\d \tint s_\al^\d + a_0 \tint s_\al + \tint s_\al^\d \tint s_\al. \label{b7} \ea
By using properties (\ref{b2}) to average Eqs. (\ref{b5}) -- (\ref{b7}), one finds that
\ba  \<a_\ta^2\> &= &\<a_0^2\> (1 + \al\ta), \label{b8} \\
\<a_\ta^\d a_\ta\> &= &\<a_0^\d a_0\> (1 + \al\ta) + \<\tint s_\al\tint s_\al^\d\>, \label{b9} \\
\<a_\ta a_\ta^\d\> &= &\<a_0a_0^\d\> (1 + \al\ta) + \<\tint s_\al^\d\tint s_\al\>. \label{b10} \ea
Notice that the equation for the squared moment does not contain a noise term.
It follows from Eqs. (\ref{b9}) and (\ref{b10}) that
\be \<[a_\ta,a_\ta^\d]\> = \<[a_0,a_0^\d]\> (1 + \al\ta) - \<[\tint s_\al, \tint s_\al^\d]\>, \label{b11} \ee
where $\<[a_0, a_0^\d]\> = 1$. By using properties (\ref{b3}) to evaluate the last commutator in Eq. (\ref{b11}), one finds that it equals $(\si_{\rm a} - \si_{\rm n})\ta$. Hence, $\<[a_\ta,a_\ta^\d]\> = 1$ if and only if $\si_{\rm a} - \si_{\rm n} = \al$. This constraint specifies the difference between the coefficients, but not their absolute values, so properties (\ref{b3}) can be rewritten in the forms
\be \<s_\al^\d(t)s_\al(t')\> = \al \si_\al\de(t - t'), \ \ \<s_\al(t)s_\al^\d(t')\> = \al(\si_\al + 1)\de(t - t'), \label{b12} \ee
where the (dimensionless) coefficient $\si_\al$ remains to be determined. A similar analysis for $\bt \neq 0$ ($\tint s_\al^\d \rightarrow \tint s_\bt$) produces similar expectation values ($\al \rightarrow \bt$). Properties (\ref{b12}) are the continuous limits of those one would obtain by considering a sequence of short gain (loss) processes, each of which has its own auxiliary mode.

One can deduce the significance of the $\si$ coefficients by considering examples. First, consider parametric amplification \cite{cav82,mck05}, in which signal and idler photons are produced in pairs. The generation of a signal photon can be stimulated by a signal photon or an idler photon, at a rate that is proportional to the number of signal and idler photons. It also can be stimulated by the quantum fluctuations of the idler, in which case the process is called spontaneous emission. If the idler is in a vacuum state, there are no idler photons, so only spontaneous emission can occur. However, if the idler is in a thermal state, idler photons are present that can stimulate signal-photon emission. According to the second of Eqs. (\ref{b12}), the number of signal photons generated in the time $\ta$ is $\al(\si_\al + 1)\ta$. Hence, $\si_\al$ is the mean number of idler photons.

Second, consider the beam-splitting model of loss \cite{lou00,mck05}. Signal photons are converted into loss-mode photons at a rate that is proportional to the number of signal photons. Likewise, loss-mode photons are converted into signal photons at a rate that is proportional to the number of loss-mode photons. If the loss mode is in a vacuum state, there are no loss-mode photons, so no signal photons are produced. However, if the idler is in a thermal state, loss-mode photons are present and can be converted. According to the first of Eqs. (\ref{b12}), the number of signal photons generated in the time $\ta$ is $\bt\si_\bt\ta$. Hence, $\si_\bt$ is the mean number of loss-mode photons.

Third, consider a signal mode that is in a thermal state initially (before the pump is turned on). If only loss were present, the number of signal photons would decrease. Equilibrium is maintained by the interaction of the signal mode with a reservoir of other modes \cite{lou73}. Signal photons can be converted to reservoir photons, but so also can reservoir photons be converted to signal photons. This interaction allows the signal and reservoir to reach an equilibrium, in which the number of signal photons equals the number of reservoir photons at the signal frequency. One can model this process by setting $\si_\bt = \<n_r\> = 1/(e^{\hb\om/kT} - 1)$. Then, according to Eq. (\ref{3.4}), $\<n\> \rightarrow \si_b/\bt = \si_\bt = \<n_r\>$. In this example, signal loss (dissipation) and noise-driven signal fluctuations are different aspects of the same interaction. This relationship is called the fluctuation--dissipation theorem.

With the terms in Eq. (\ref{b1}) specified completely, one can derive equations for the amplitude and quadrature moments. It follows from Eqs. (\ref{b1}) and (\ref{b2}) that
\be d_t \<a\> = (\al - \bt)\<a\>/2, \label{b21} \ee
where $\<\ \>$ denotes an expectation value. Gain increases the amplitude mean, whereas loss decreases it. The quadrature operator $q(\ph) = (a^\d e^{i\ph} + a e^{-i\ph})/2^{1/2}$, where $\ph$ is the phase of the local oscillator. By combining this definition with Eq. (\ref{b21}), one obtains the quadrature-mean equation
\be d_t \<q(\ph)\> = (\al - \bt)\<q(\ph)\>/2. \label{b22} \ee
Gain and loss change the quadrature mean, in ways that are insensitive to the local-oscillator phase.

It follows from Eqs. (\ref{b8}) -- (\ref{b10}) and (\ref{b12}) that
\ba  d_t \<a^2\> &= &(\al - \bt)\<a^2\>, \label{b23} \\
d_t \<a^\d a\> &= &(\al - \bt)\<a^\d a\> + \al(\si_\al + 1) + \bt \si_\bt, \label{b24} \\
d_t \<aa^\d\> &= &(\al - \bt)\<aa^\d\> + \al \si_\al + \bt(\si_\bt + 1). \label{b25} \ea
The second quadrature moment
\be \<q^2(\ph)\> = \<(a^\d)^2 e^{i2\ph} + a^\d a + aa^\d + a^2 e^{-i2\ph}\>/2. \label{b26} \ee
By combining Eqs. (\ref{b23}) -- (\ref{b26}), one obtains the second-moment equation
\be d_t \<q^2(\ph)\> = (\al - \bt)\<q^2(\ph)\> + \al(\si_\al + 1/2) + \bt(\si_\bt + 1/2). \label{b27} \ee
The quadrature deviation $\de q = q - \<q\>$, so the quadrature variance $\<\de q^2\> = \<q^2\> - \<q\>^2$. By combining Eqs. (\ref{b22}) and (\ref{b27}), one obtains the quadrature-variance equation
\be d_t \<\de q^2\> = (\al - \bt)\<\de q^2\> + \al(\si_\al + 1/2) + \bt(\si_\bt + 1/2). \label{b28} \ee
Gain and loss also change the quadrature variance in phase-insensitive ways. The last two terms in Eq. (\ref{b28}) represent the fluctuations contributed by the (implicit) idler and loss modes. Notice that the vacuum term in the number-mean equation (\ref{b24}) is $\al$, whereas the vacuum terms in the quadrature-variance equation are $\al/2$ and $\bt/2$.

In the semi-classical model of gain and loss, one replaces the mode operator $a$ by the mode amplitude $A$, which has both deterministic and random components. It follows from Eq. (\ref{2.1}) that
\be d_t \<A\> = (\al - \bt)\<A\>/2. \label{b29} \ee
Equation (\ref{b29}) is consistent with Eq. (\ref{b21}). The semi-classical quadrature $Q$ is defined in the same way as the quantum one. By combining this definition with Eq. (\ref{b29}), one obtains the quadrature-mean equation
\be d_t \<Q(\ph)\> = (\al - \bt)\<Q(\ph)\>/2. \label{b30} \ee
Equation (\ref{b30}) is consistent with Eq. (\ref{b22}). The semi-classical derivation of the quadrature-variance equation is similar to the quantum derivation. The only differences are that the 1s in Eqs. (\ref{b24}) and (\ref{b25}) are absent, and the second and third terms on the right side of Eq. (\ref{b26}) contribute equally. Hence,
\be d_t \<\de Q^2\> = (\al - \bt)\<\de Q^2\> + \sb_\al + \sb_\bt, \label{b31} \ee
where $\sb_\al$ and $\sb_\bt$ are the semi-classical strength coefficients. One can reconcile Eq. (\ref{b31}) with Eq. (\ref{b28}) by setting $\sb_\al = \al(\si_\al + 1/2)$ and $\sb_\bt = \bt(\si_\bt + 1/2)$. These replacements are consistent with the definition of the source strength in Eq. (\ref{2.2}). Thus, the semi-classical model of gain and loss predicts the changes in quadrature mean and variance accurately.
(Further analysis shows that it overestimates the number means by 1/2 and the number variances by 1/4.)

\section*{Appendix C: van der Pol oscillator}

The textbook van der Pol (vdP) equation \cite{jor07} is
\be d_{tt}^2 X - 2\ep(1 - X^2)d_t X + X = 0, \label{c1} \ee
where $X$ is the dependent variable (displacement), $d_t$ is a time derivative and $\ep > 0$ is a parameter (not the enhancement factor). Equation (\ref{c1}) describes an oscillator with a linear restoring force, a linear negative-damping (driving) term and a nonlinear damping term. The net-damping rate is negative if $|X| < 1$ and positive if $|X| > 1$. In this appendix, the relation between Eqs. (\ref{4.4}) and (\ref{c1}) is determined.

Suppose that the damping terms are small, so that perturbation theory can be used to model their effects. Let $X = X_0 + \ep X_1$ be the perturbed displacement and let $t_0 = t$ and $t_1 =\ep t$ be fast and slow time variables, respectively. These variables are treated as independent, in which case the total time derivative $d/dt = d/dt_0 + \ep d/dt_1$, which one can abbreviate as $D_0 + \ep D_1$. By making these substitutions in Eq. (\ref{c1}), and retaining terms of order 1 and $\ep$, one obtains the perturbed equation
\be (D_0^2 + 2\ep D_0D_1)(X_0 + \ep X_1) - 2\ep(1 - X_0^2)D_0X_0 + (X_0 + \ep X_1) = 0. \label{c2} \ee

By collecting terms of order 1, one obtains the zeroth-order equation
\be D_0^2 X_0 + X_0 = 0. \label{c3} \ee
Equation (\ref{c3}) has the periodic solution
\be X_0(t_0,t_1) = A_0(t_1) e^{-it_0} + c.c., \label{c4} \ee
where the slow-time evolution of the zeroth-order amplitude remains to be determined.

By collecting terms of order $\ep$, one obtains the first-order equation
\be D_0^2 X_1 + X_1 = -2D_0D_1 X_0 + 2(1 - X_0^2)D_0X_0. \label{c5} \ee
Equation (\ref{c5}) describes a driven first-order displacement. If the terms on the right side have a resonant component (which is proportional to $e^{-it_0}$), the displacement will grow without bound. This behavior is unphysical, so the resonant component of the right side must be zero. The resonant components of the first and second terms are $2iD_1A_0e^{-it_0}$ and $-2iA_0e^{-it_0}$, respectively. The third term,
\be -2X_0^2D_0X_0 = -2[A_0^2e^{-2it_0} + 2|A_0|^2 + (A_0^*)^2e^{2it_0}] (-iA_0e^{-it_0} + iA_0^*e^{it_0}), \label{c6} \ee
has the resonant component $2i|A_0|^2A_0e^{-it_0}$. By combining these results, one obtains the slow-time amplitude equation
\be D_1A_0 = (1 - |A_0|^2)A_0, \label{c7} \ee
which has the same form as Eq. (\ref{4.4}). Thus, Eq. (\ref{4.4}) is the weak-damping limit of Eq. (\ref{c1}). The reduced equation describes the slow growth and saturation of the complex amplitude. It describes the evolution of a variety of physical systems, which have stable nonlinear equilibria. Notice that the phase of $A_0$ is unconstrained, so in the presence of complex amplitude noise, it can change significantly.

\section*{Appendix D: Integral of colored noise}

In \cite{mck21}, we derived formulas for the temporal correlation and frequency spectrum of an undamped oscillator driven by white noise. Such an oscillator is governed by the amplitude equation
\be d_t B = R(t), \label{d1} \ee
where the source function has properties (\ref{2.2}). In this appendix, analogous formulas are derived for an undamped oscillator driven by colored noise, which has the correlation
\be \<R^*(t)R(t')\> = \ka e^{-\nu|t - t'|}. \label{d2} \ee
These formulas are relevant to the study of phase fluctuations, which are driven by a combination of white and colored noise [Eq. (\ref{4.6})].
Equation (\ref{d1}) has the solution
\be B(t) = \int_0^t R(s) ds. \label{d3} \ee
In the same way that Brownian motion is the integral of white noise, solution (\ref{d3}) is the integral of colored noise.

It follows from the first of Eqs. (\ref{2.2}) and Eq. (\ref{d3}) that the amplitude mean $\<B(t)\> = 0$.
By combining Eqs. (\ref{d2}) and (\ref{d3}), one finds that the amplitude variance
\ba \<|B(t)|^2\> &= &\int_0^t \int_0^t \ka e^{-\nu|s - s'|} dsds' \nonumber \\
&= &2\ka \int_0^t \int_0^s e^{-\nu(s - s')} ds'ds \nonumber \\
&= &(2\ka/\nu) \int_0^t (1 - e^{-\nu s}) ds \nonumber \\
&= &(2\ka/\nu)[t - (1 - e^{-\nu t})/\nu]. \label{d4} \ea
For times that are longer than the correlation time $1/\nu$, the variance produced by colored noise grows linearly with time (as does the variance produced by white noise). In Sec. 4, we rewrote the right side of Eq. (\ref{d2}) as $\si\de_e(\ta)$, where $\si = 2\ka/\nu$ is a strength coefficient and $\de_e(\ta) = (\nu/2)e^{-\nu|\ta|}$ is an effective $\de$-function. Then we replaced the effective $\de$-function by an actual one and obtained the variance $\si t$. This approach produced the leading term in Eq. (\ref{d4}). Furthermore, the correction term $-\si/\nu$ is constant, so it affects the complex-amplitude correlation slightly, but does not affect the spectrum at all [Eqs. (\ref{5.4}) -- (\ref{5.7})]. In this context, the white-noise approximation is very accurate.

For $t' > t$, the temporal correlation
\ba \<B^*(t)B(t')\> &= &\int_0^t \int_0^{t'} \ka e^{-\nu|s - s'|} ds'ds \nonumber \\
&= &\ka \int_0^t \int_0^s e^{-\nu(s - s')} ds'ds + \ka\int_0^t \int_s^{t'} e^{-\nu(s' - s)} ds'ds \nonumber \\
&= &(\ka/\nu) \int_0^t (1 - e^{-\nu s}) ds + (\ka/\nu) \int_0^t [1 - e^{-\nu(t' - s)}] ds \nonumber \\
&= &(\ka/\nu) \{t - (1 - e^{-\nu t})/\nu + t - [e^{-\nu(t' - t)} - e^{-\nu t'}]/\nu\} \nonumber \\
&= &(\ka/\nu) \{2t - [1 - e^{-\nu t} - e^{-\nu t'} +  e^{-\nu(t' - t)}]/\nu\}. \label{d5} \ea
Because the correlation is real, it is a symmetric function of $t$ and $t'$, so for $t > t'$, the first term on the right side of Eq. (\ref{d5}) is proportional to $t'$ and the exponent of the last term is $-\nu(t - t')$. Thus, the correlation depends on the minimal time $\min(t,t')$ and the time difference $|t - t'|$. Notice that for times longer than the correlation time, the correlation produced by colored noise is proportional to the minimal time (as is the correlation produced by white noise). In this context, the white-noise approximation is accurate.

By combining Eqs. (\ref{2.6}) and (\ref{d5}), one obtains the spectrum equation
\ba \<|B(\om)|^2\> &= &(\ka/\nu) \int_0^T \int_0^T [2\min(t,t') \nonumber \\
&&-\ (1 - e^{-\nu t} - e^{-\nu t'} + e^{-\nu|t - t'|})/\nu]e^{i\om(t' - t)} dtdt'. \label{d6} \ea
The first integral in Eq. (\ref{d6}) is the white-noise integral that was done in \cite{mck21}. The result is
\be I_1 = 4T[1 - \sinc(\om T)]/\om^2. \label{d7} \ee
Notice that $I_1 \rightarrow 2T^3/3$ as $\om \rightarrow 0$ and $4T/\om^2$ as $\om \rightarrow \infty$.

The second integrand is a separable function of $t$ and $t'$, so
\be I_2 = \frac{(e^{-i\om T} - 1)(e^{i\om T} - 1)}{(-i\om)(i\om)\nu} = \frac{2[1 - \cos(\om T)}{\nu\om^2}. \label{d8} \ee
Notice that $I_2 \rightarrow T^2/\nu$ as $\om \rightarrow 0$ and is proportional to $2/\nu\om^2$ for high frequencies.

The third integrand does not depend on $t'$, so
\be I_3 = \Biggl[\frac{e^{i\om T} - 1}{i\om}\Biggr] \Biggl[\frac{1 - e^{-(\nu + i\om)T}}{(\nu + i\om)\nu}\Biggr]. \label{d9} \ee
The fourth integrand does not depend on $t$, so
\be I_4 = \Biggl[\frac{e^{-i\om T} - 1}{-i\om}\Biggr] \Biggl[\frac{1 - e^{-(\nu - i\om)T}}{(\nu - i\om)\nu}\Biggr]. \label{d10} \ee
By combining these results, one obtains the joint contribution
\be I_3 + I_4 = \frac{2\sin(\om T)}{\om(\nu^2 + \om^2)} + \frac{2[1 - \cos(\om T)]} {\nu(\nu^2 + \om^2)}, \label{d11} \ee
where the exponentially small terms were neglected. Notice that $I_3 + I_4 \rightarrow 2T/\nu^2$ as $\om \rightarrow 0$ and is proportional to $2/\nu\om^2$ for high frequencies.

The fifth integral is split into two parts, the first of which is
\ba &&\frac{1}{\nu} \int_0^T \int_0^t e^{-(\nu + i\om)(t - t')} dt'dt \nonumber \\
&= &\int_0^T \frac{1 - e^{-(\nu + i\om)t}}{\nu(\nu + i\om)} dt \nonumber \\
&= &\frac{T}{\nu(\nu + i\om)} - \frac{1 - e^{-(\nu + i\om)T}}{\nu(\nu + i\om)^2}, \label{d12} \ea
and the second of which is
\ba &&\frac{1}{\nu} \int_0^T \int_t^T e^{-(\nu - i\om)(t' - t)} dt'dt \nonumber \\
&= &\int_0^T \frac{1 - e^{-(\nu - i\om)(T - t)}}{\nu(\nu - i\om)} dt \nonumber \\
&= &\frac{T}{\nu(\nu - i\om)} - \frac{1 - e^{-(\nu - i\om)T}}{\nu(\nu - i\om)^2}. \label{d13} \ea
By combining these results, one finds that
\be I_5 = \frac{2T}{\nu^2 + \om^2} - \frac{2(\nu^2 - \om^2)}{\nu(\nu^2 + \om^2)^2}, \label{d14} \ee
where the exponentially small terms were neglected. Notice that $I_5 \rightarrow 2T/\nu^2$ as $\om \rightarrow 0$ and $2T/\om^2$ as $\om \rightarrow \infty$.

It is convenient to write the frequency spectrum in terms of the coefficient $\si = 2\ka/\nu$, which, as explained after Eq. (\ref{d4}), is the strength of the equivalent white-noise source. By multiplying the preceding integrals by $\ka/\nu = \si/2$, one obtains the spectrum
\ba \<|B(\om)|^2\> &= &\frac{2\si T[1 - \sinc(\om T)]}{\om^2} - \frac{\si[1 - \cos(\om T)]}{\nu\om^2} + \frac{\si\sin(\om T)}{\om(\nu^2 + \om^2)} \nonumber \\
&&+\ \frac{\si[1 - \cos(\om T)]}{\nu(\nu^2 + \om^2)} - \frac{\si T}{\nu^2 + \om^2} + \frac{\si(\nu^2 - \om^2)}{\nu(\nu^2 + \om^2)^2} \nonumber \\
&= &\frac{2\si T[1 - \sinc(\om T)]}{\om^2} - \frac{\si T[1 - \sinc(\om T)]}{\nu^2 + \om^2} \nonumber \\
&&-\ \frac{\si\nu[1 - \cos(\om T)]}{\om^2(\nu^2 + \om^2)} + \frac{\si(\nu^2 - \om^2)} {\nu(\nu^2 + \om^2)^2}. \label{d15} \ea
As $\om \rightarrow 0$, the first term in Eq. (\ref{d15}) tends to $\si T^3/3$, and the other terms tend to 0, $-\si T^2/2\nu$ and $\si/\nu^3$. As $\om \rightarrow \infty$, the first term tends to $2\si T/\om^2$ and the other terms tend to $-\si T/\om^2$, $-\si\nu/\om^4$ and $-\si/\nu\om^2$. Overall, the first two terms are the most importantant. The first term is the white-noise contribution to the spectrum and the second is the colored-noise correction. For low frequencies, the spectrum is consistent with Eq. (\ref{4.15}), whereas for high frequencies the colored-noise correction reduces the spectrum by a factor of 2. In this context, the white-noise approximation is reasonable.

\section*{Data availability}

No data were generated or analyzed in this research.

\section*{Disclosure}

The authors declare that there is no conflict of interest.

\section*{Funding} TJS and ASH were supported by the Natural Sciences and Engineering Research Council of Canada.

\end{document}